\documentclass[
 reprint,
 aps,
 superscriptaddress
]{revtex4-1}

\usepackage{graphicx}

\begin{document}

\title{A Non-stoichiometric Universality in Microbubble-Polyelectrolyte Complexation
}
\author{Hiroshi Frusawa}
\email{frusawa.hiroshi@kochi-tech.ac.jp}
\author{Ryohei Yoshida}
\affiliation{Institute for Nanotechnology, Kochi University of Technology, Tosa-Yamada, Kochi 782-8502, Japan.}


\begin{abstract}
We investigated the fundamental electrical properties of microbubbles (MBs) that are directly encapsulated by the addition of oppositely charged polyelectrolytes (PEs). 
Charge-reversal of the MB-PE complex particles has been observed by the microscopic electrophoresis method, revealing unusual overcharging behaviors in MB-PE complex solutions, as follows.
The critical concentrations of cationic PEs added for overcharging were not only independent of their chain lengths and molecular species, but also much larger than stoichiometric neutralization points.
Thus, we provide a theoretical sketch that considers the adsorption-desorption kinetics of small anions on the surface of genuine microbubbles, which can explain the inefficient charge-reversal.
\end{abstract}

\maketitle

\section{Introduction}

Microbubbles (MBs) designate colloidal bubbles with a diameter ranging from 0.1 to 100 micrometers.
Various techniques for creating micron- and submicron-bubbles have been found, such as sonication or high-shear emulsification, inkjet printing, and microfluidic processing \cite{stride}.
Because each MB has a large Laplace pressure driving bubble dissolution, the interior gas tends to be leaked into aqueous solutions \cite{duncan}.
Coating the gas core by a shell increases the lifetime of individual bubbles.
These shells have been fabricated using a variety of materials, such as proteins, lipids  and nanoparticles. Coating the gas particles by various shells slows down gas diffusion, which results in stability that lasts for a few months \cite{borden}.

Low gas permeability, leading to efficient trapping of encapsulated gas inside solutions, is of great interest not only in terms of green-technological applications including fuel cell components and industrial food applications like flavor additives, but also in many biomedical applications \cite{borden2, gene}.
Because MBs efficiently absorb ultrasound waves in vivo, the encapsulated MBs are applicable for ultrasound contrast agents, targeted drug delivery, and gene therapy \cite{gene}.
One of the most promising shells is the polyelectrolyte (PE) multilayer formed by a layer-by-layer method that sequentially deposits oppositely charged PEs onto the gas core material \cite{angew1, angew2}.
Moreover, these polymeric shells impart good biocompatibility and biodegradability  and may provide additional functionality, such as molecular recognition.
In particular, the cationic shells of the PE multilayer are useful for DNA vehicles in gene therapy because nucleic acids can electrostatically couple directly to the surface of MB capsules.
The multilayers also have the potential advantage of increasing the overall loading capacity of MB-surfaces due to the electrostatic sandwiching of DNA between cationic layers.
We have obtained positive results of MBs with DNA shells that enhance in vivo transfection in gene therapy, using ultrasound-targeted destruction \cite{lang, borden3}.

The layer-by-layer method essentially makes use of the charge reversal phenomena where colloidal particles are overcharged by adding oppositely charged PEs;
however, the first layer covering the surface of genuine MBs seems to be an exception.
Previous encapsulations \cite{angew1, angew2, lang, borden3} have typically been prepared with a lipid or protein shell substrate on which the first layer of PE chains are attached instead of direct deposition onto a genuine gas core.
There is a recent report \cite{daiguji} that forms the first PE layer without the use of other materials, though this trial involves several steps, such as the introduction of $\mathrm{CO_2}$ MBs.
These results imply difficulty in implementing the charge reversal of unmodified MBs that carry anions due to the adsorption of hydroxyl ions ($\mathrm{OH}^{-1}$) on the gas surfaces \cite{takahashi}.
These surfaces are similar to oil-water interfaces \cite{velev} with no salt added, though such fundamental issues have been beyond the scope of MB research other than an investigation \cite{frusawa} by one of the present authors.
Here, we investigate in more detail the charge-reversal phenomena of MBs using the microscopic electrophoresis method.

\section{Experimental Section}

We used an MB generator (OM4-GP-040, Aura Tech) for the production of submicron air bubbles from deionized water.
The salt-free MB solutions were in the pH range of 6.5 to 7.0 \cite{frusawa}.
The cationic PEs that were added to the genuine MB solutions were poly-L-lysine (PLL, molecular weight (Mw) 70,000-150,000 , Wako) and two types of poly(allylamine hydrochloride) (PAH): long PAH (Mw 56,000, Sigma) and short PAH (Mw 17,000, Sigma).
We also used polyethylene glycol (Mw 5,000-7,000, Fluka) as a neutral polymer and poly(4-ammonium styrene sulfonic acid) (Mw 200,000, Sigma) as an anionic PE.
To eliminate aggregates or dusts, we passed the cationic solutions through a 200-nm pore-size filter (Minisart, Sartorius) before mixing.
The pH of the MB-PE complex solutions ranged from 7.2 to 7.8.
For comparison, we also measured the electrophoretic velocity of silica particles (SPs, Duke Scientific) with a nominal diameter of 1 $\mu\mathrm{m}$ as a reference for the anionic colloids.
Although the SP solutions were stable with a pH of 7.0, the SP solutions with PEs added had a pH range of 7.5 to 8.0.

We detected gas particles undergoing Brownian motion and electrophoresis at 20Ž via dark-field microscopy using a zeta potential analyzer (Zeecom, Microtech).
Video observations of the migrating particles enabled individual velocities to be measured by tracking the particles, which has been referred to as the microscopic electrophoresis method.
We used a rectangular cell of 1-cm height ($2h=10$ mm), 0.75-mm depth ($2d=0.75$ mm), and 9-cm length equal to the electrode gap length.
The inherent electrophoretic drift velocity is obtained by fitting the theoretical velocity distribution to the measured velocity as a sum of electrophoretic and electroosmotic velocity profiles inside a cell.
The theoretical distribution function for the rectangular housing is of the form \cite{palberg},
   \begin{equation}
   v(x)=v_{\mathrm{eo}}\left[
   1-3\left\{
   \frac{1-(x/d)^2}{2-384d/(\pi^5h)}
   \right\}
   \right],
   \end{equation}
where $v_{\mathrm{eo}}$ denotes the electroosmotic velocity in the vicinity of the cell walls, $\pm d$ the extension in the direction of depth, and $2h$ the height.

\section{Results}
\subsection{Mean size of microbubbles as Brownian particles}

We evaluated the mean hydrodynamic radius of MBs that undergo Brownian motion using dark-field microscopy, as follows.
The video analysis provides the mean square displacement (MSD), $\left<{\bf r}^2(t)\right>$, of tracer particles, which is proportional to a time interval $t$ when the particle displacement is ${\bf r}(t)$:
   \begin{equation}
   \left<{\bf r}^2(t)\right>=2dDt,
   \label{msd}
   \end{equation}
where $D$ denotes the diffusion constant. We adopted $d=2$ as the spatial dimension because the displacement vector ${\bf r}(t)$ lies in a two-dimensional plane when measuring the MSD of the Brownian trajectories of the tracer particles in the videos.
Furthermore, we used the usual Stokes-Einstein relation,
   \begin{equation}
   D=\frac{k_BT}{6\pi\eta a},
   \label{se}
   \end{equation}
in the diffusion processes of MBs, with $k_BT$ being the thermal energy, $\eta$ the water viscosity, and $a$ the hydrodynamic radius. 
From eqs. (\ref{msd}) and (\ref{se}), the MSD measurements were found to give mean diameters, $2a$, of the MBs in various experimental conditions.
\begin{figure}[tbh]
\begin{center}
\includegraphics[width=7cm]{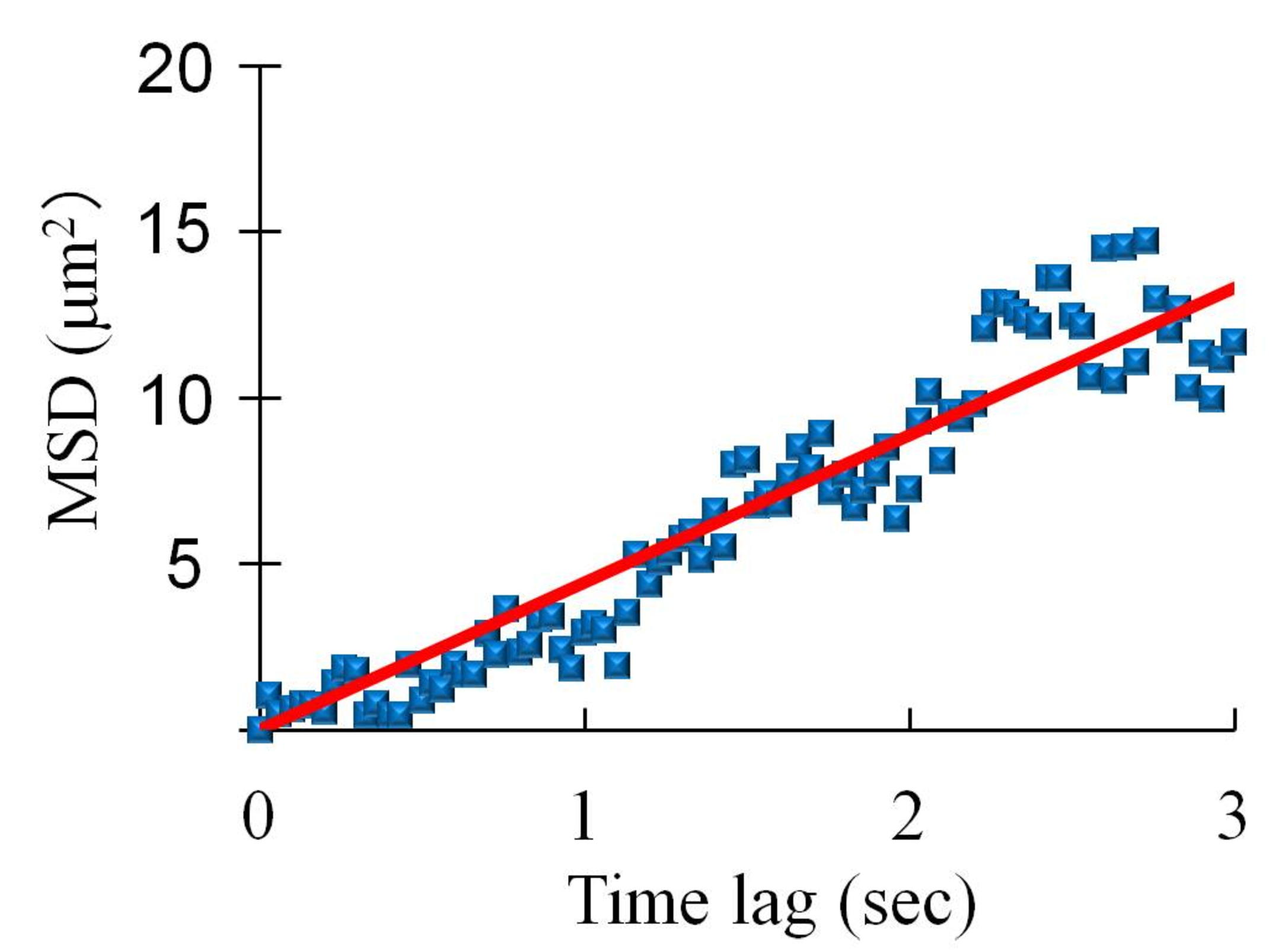}
\caption{(Color online) The mean square displacement (MSD) of Brownian particles in microbubble solutions increases as time interval is larger. The solid line delineates eq. (2) that is fitted to the obtained data.}
\label{f1}
\end{center}
\end{figure}

Figure 1 shows an MSD  versus time lag for genuine MBs in water at 20 Ž .
We evaluated from eqs. (\ref{msd}) and (\ref{se}) that the mean-hydrodynamic diameter of the observed Brownian particles is $2a\approx400$ nm.
The MSD measurements also verified that most MB-PE complexes remain approximately 400 nm in diameter \cite{frusawa}, except for a few aggregates that almost entirely isoelectric solutions are likely to include per micrograph.
The dark-field microscopy, however, facilitates distinguishing these particles from the other normal particles because the brightness of scattered light by the aggregates is apparently different from the others.
Thus, we avoided tracking the aggregated particles in electrophoretic measurements.

\subsection{Effective charge density on microbubble surfaces}

Figure 2 demonstrates the linear relationship between the applied electric field and the inherent electrophoretic mobility that is extracted from the analysis of positionally dependent velocities of both genuine MBs and SPs as described in section 2.
The slopes of the fit lines in Fig. 2 correspond to the electrophoretic mobilities of MBs and SPs, indicating that the MB-mobility is close to that of the SPs.
That is, Fig. 2 implies that the effective charge number $Z^*_{\mathrm{MB}}$ per uncoated MB is not far from that of SP ($Z^*_{\mathrm{SP}}$), though salt-free MB solutions have no charged molecules, other than dissociated water molecules.
In what follows, we evaluate $Z^*_{\mathrm{MB}}$ and $Z^*_{\mathrm{SP}}$ quantitatively using the H\"uckel formula \cite{huckel} of electrophoretic mobility.
\begin{figure}[tbh]
\begin{center}
\includegraphics[width=8cm]{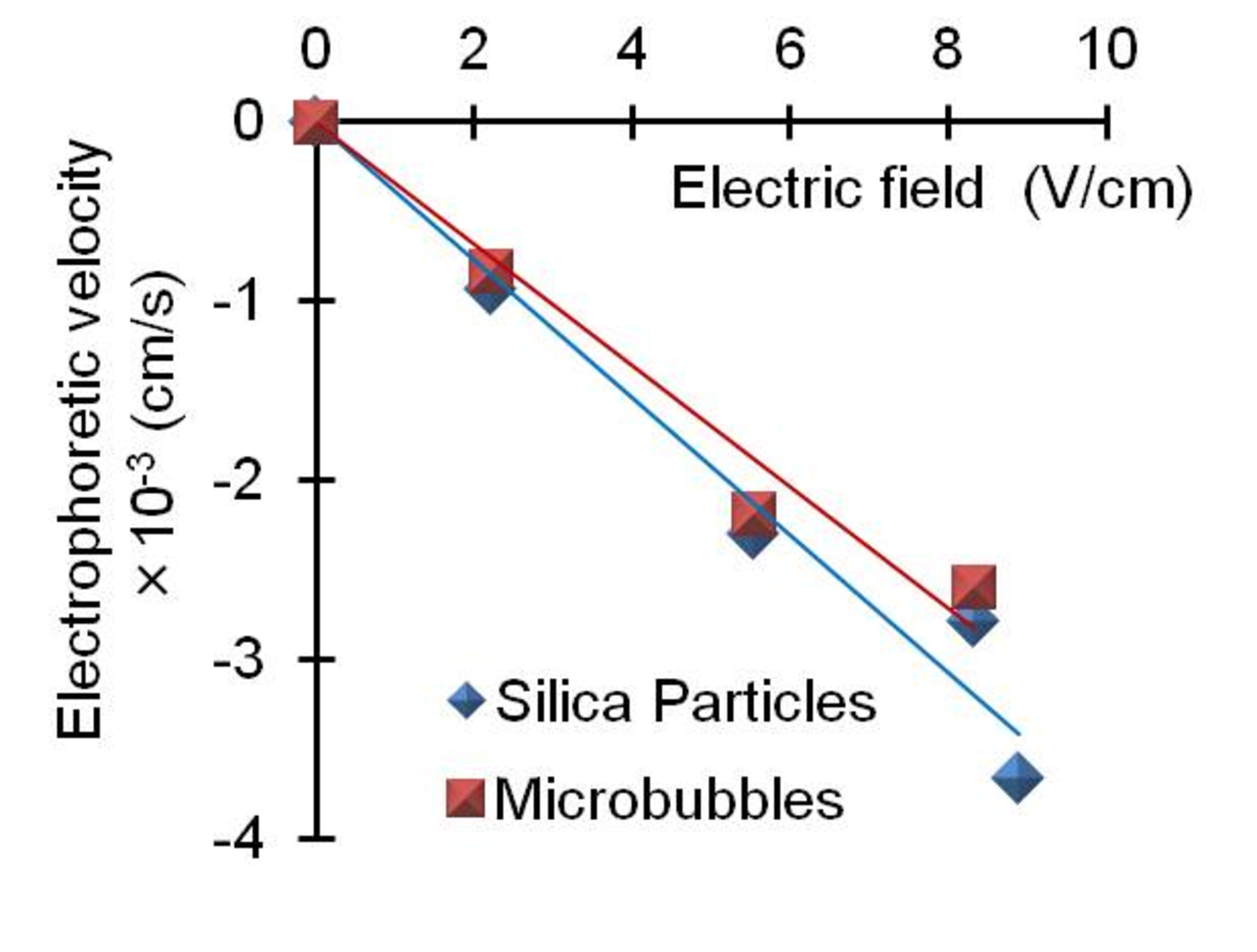}
\caption{(Color online) The mean electrophoretic velocities of microbubbles dependent on applied electric fields, which are represented by fitting a straight line with a slope that corresponds to the mobility. A comparison of the data for microbubbles and silica particles is also shown, demonstrating their similarity.}
\label{f2}
\end{center}
\end{figure}

The Debye-H\"uckel screening length $\kappa^{-1}$ is generally defined by $\kappa^2=4\pi l_BI$ where ionic strength $I$ is due to free small ions and the Bjerrum length $l_B$ is given by $l_B=e^2/(4\pi\epsilon k_BT)$, with $e$ indicating the elementary charge and $\epsilon$ indicating the permittivity of the solvent.
In the present study, we use $l_B=0.73$ nm, which is calculated for water at 20 Ž.
In the case of salt-free MB solutions, the ionic strength is given by $I=q_{\mathrm{w}}$, where $q_{\mathrm{w}}$ represents the sum of charge density arising from dissociated water molecules.
We adopt $q_w=2\times10^{-7}$ M, for simplicity, in the above pH range (see section 2), yielding $\kappa\approx 0.2$.

Recent simulation results demonstrate that the following H\"uckel formula is valid for $\kappa a<1$ \cite{huckel}: $\mu_{\mathrm{red}}=Z^*_{\mathrm{MB}}(1+\kappa a)^{-1}(l_B/a)$,
where the reduced dimensionless mobility is defined by $\mu_{\mathrm{red}}=\left\{3e\eta/(2\epsilon k_BT)\right\}\mu$ using the measured one $\mu$.
Using this relation, we evaluate $Z^*_{\mathrm{MB}}\approx 9\times10^2$ for MBs, and $Z^*_{\mathrm{SP}}\approx3.5\times 10^3$ for SPs.
Because the SP-diameter is approximately 2.5 times larger than that of MBs, a comparison in terms of the effective charge densities of MBs and SPs, $\sigma^*_{\mathrm{MB}}$ and $\sigma^*_{\mathrm{SP}}$, would be more relevant.
The obtained effective charges were $\sigma^*_{\mathrm{MB}}\approx -1.7\times10^{-3}e\,\mathrm{C}/\mathrm{nm}^2$ and $\sigma^*_{\mathrm{SP}}\approx-1.1\times10^{-3}e\,\mathrm{C}/\mathrm{nm}^2$, revealing comparable charge densities of genuine MBs and SPs, respectively.

\subsection{Charge reversal of microbubble-polyelectrolyte complexes}

Figure 3 displays variations in the slope of the linear relationships between the applied electric field and the inherent electrophoretic mobility of various MB-PLL complexes.
As the monomer concentration of PLL, $C_{\mathrm{LL}}$, is increased, the slope, or the electrophoretic mobility increases, and the sign of the mobility is reversed beyond a critical PLL concentration $C_{\mathrm{LL}}^*$ as determined below.
Figure 3 illustrates that the electrophoretic mobility of MB-PLL complexes is negligible when PLL with a monomer concentration of $C_{\mathrm{LL}}=3\,\mu \mathrm{M}(>C^*_{\mathrm{LL}})$ is added.
We have also performed electrophoresis experiments of SPs with long PAHs added and have determined the critical monomer density $D^*_{\mathrm{LAH}}$ as a reference by adjusting the mean total number, which is evaluated by counting the particles per micrograph, with that of MBs.
Nevertheless, the obtained density was $D^*_{\mathrm{LAH}}\approx0.07\,\mu \mathrm{M}$ for the SPs, which is much smaller than the aforementioned MB density of $C_{\mathrm{LL}}=3\,\mu \mathrm{M}$.
\begin{figure}[tbh]
\begin{center}
\includegraphics[width=11cm]{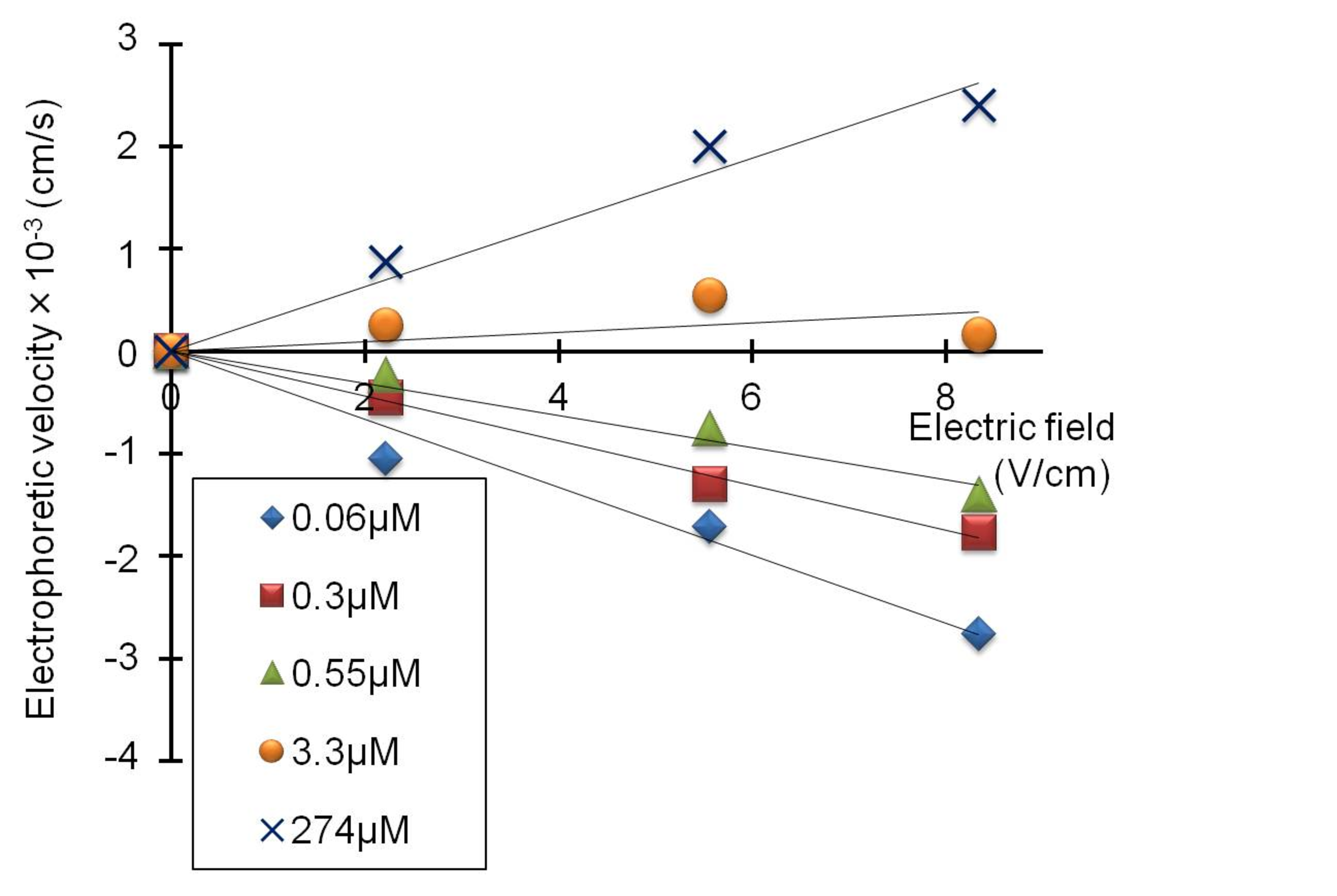}
\caption{(Color online) The data for various monomer concentrations of added PLL and straight fit lines demonstrating a variation of slopes in the graph of electrophoretic velocity vs. electric field. The sign change from negative to positive slopes reflects the overcharging of microbubbles due to the addition of PLL.}
\label{f3}
\end{center}
\end{figure}

The following consistency of neutralized SP-PAH complexes with the stoichiometry conversely provides clear evidence of the extraordinary concentration $C_{\mathrm{LL}}^*$ far beyond the stoichiometric point.
In the evaluation of the bare charge number $Z_{\mathrm{SP}}$ that each SP carries, the following two aspects are supposed: 
the first is that the stoichiometry at the isoelectric point \cite{conventional} can apply to the SP systems, and the other is to regard the obtained charge density not as effective but as a bare value assuming that the complexation results in the release of small counterions.
We thus have the stoichiometric relation, $C^*_{\mathrm{LL}}+Z_{\mathrm{SP}}C_{\mathrm{SP}}=0$, yielding the bare charge density $\sigma_{\mathrm{SP}}$ on SPs that is given by $\sigma_{\mathrm{SP}}=|-Z_{\mathrm{SP}}e/(4\pi a^2)|\approx5e\,\mathrm{C}/\mathrm{nm}^2$ where $a= 500$ nm and $e$ denotes the elementary charge.
The present value ($\sigma_{\mathrm{SP}}/e\approx5$) per one -nanometer area is much larger than the effective one evaluated above but agrees with the number of silanol groups per $\mathrm{nm}^2$ reported in the literature \cite{silanol}. This calculated value indicates that what is compensated for with the total positive charge due to cationic PEs is not the effective negative charge dressed with a cloud of counterions, but the bare anionic charge that MBs inherently carry \cite{conventional, dobrynin}.
The simple use of the above electrical neutrality condition in the MB-PLL system gives a surface charge density of $-900e\,\mathrm{C}/\mathrm{nm}^2$, or 900 groups of ions per 1 $\mathrm{nm}^2$, which is unreasonable.
\begin{figure*}[tbh]
\begin{center}
\includegraphics[width=16cm]{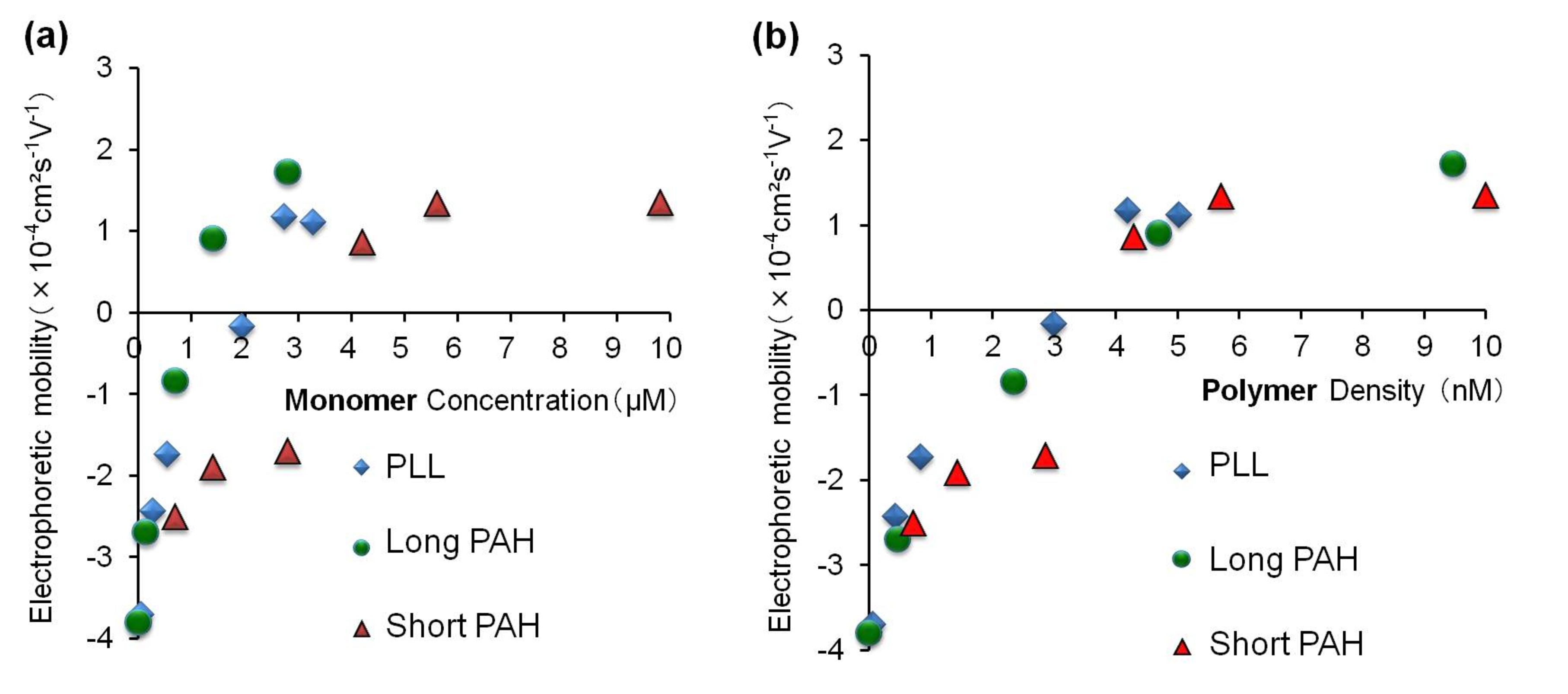}
\caption{(Color online) (a) The monomer-concentration dependences of electrophoretic mobilities when three types of cationic polyelectrolytes (PLL, long-PAH, and short-PAH) are added. Unlike conventional results of charge reversal phenomena in colloid-polyelectrolyte complexes, it is apparent that no universal curve can be fitted to these data. (b) The electrophoretic mobilities as a function of added polyelectrolyte concentrations are plotted for the same species, revealing that the sign change of the mobility occurs at a similar critical concentration of cationic polyelectrolytes.}
\label{f4}
\end{center}
\end{figure*}

Figure 4(a) depicts the dependences of electrophoretic mobilities on monomer concentrations in long-and short-PAH solutions, as well as on $C_{\mathrm{LL}}$.
Figure 4(a) also illustrates that different PEs have distinct isoelectric points where the electrophoretic mobilities vanish as a result of charge compensation.
Previous experiments of charge reversal phenomena, on the other hand, have found that the stoichiometric neutrality universally represents isoelectricity in a variety of colloidal systems \cite{conventional}.
Because an identical charge density due to the addition of oppositely charged PEs is required for the charge compensation of colloidal suspensions with equal charge density, a universal curve as a function of monomer density emerges by collecting electrophoretic mobilities of previous experiments for various polyelectrolytes with different chain lengths and/or constituent molecules.
In contrast, Fig. 4(a) exhibits that the universality principle is invalid for MB-PE complexation.
Figures 3 and 4(a) reveal unusual electrical properties where MB-PE complexes are not neutralized until an excess amount of cationic PEs are added, which the stoichiometry is unable to explain.

As a control experiment for revealing the role of opposite charges, we added non-cationic PEs (neutral polymers or anionic PE chains) beyond 100 $\mu\mathrm{M}$ in monomer concentration, where the MBs had the same mobilities as salt-free MBs within errors.
This experiment  makes it clear that PEs play the essential role of attractive electrostatic interactions in the present charge-reversal. 
Figure 4(b) implements PE density, instead of the monomer concentration, as another choice of the transverse axis that determines electrophoretic mobility.
From Fig. 4(b), we see the following:
Before overcharging, the mobilities are reduced to a similar extent, and the critical PE concentrations that effectively cancel the negative charges on MBs were evaluated from Fig. 4(b) to give approximately 3 nM for the three types of cationic PEs.
The close PE concentrations for charge compensations indicate that different amounts of cationic charges on polyelectrolytes are able to equally neutralize the anionic MBs with the same charges.
In other words, there exists a unique curve that is common to three species of PEs, which seems to be one of the key features for clarifying the novel situation in overcharging MB-PE complex particles.

\section{Discussion}
\subsection{Mean-size invariance}

We have described in section 3.1 that Brownian particles detected by dark-field microscopy have a similar diameter of approximately 400 nm under all experimental conditions, implying that the aggregation between MBs is still suppressed despite adding oppositely charged PEs.
Let us consider the reason in terms of number densities of MBs and excess amount of PEs.
The MB density is independent of the added PE-concentration and is kept constant for hours not only in MB-PE complexes \cite{frusawa} but also in genuine MB solutions \cite{frusawa, ushikubo}.
Incidentally, electrophoretic measurements need to apply three different voltages for verifying the linear relationships between electrophoretic velocities and applied fields as shown in Figs. 2 and 3,
yet it takes less than a few hours to finish the measurement set of one sample.
Therefore, this stability assures invariant MB or MB-PE systems during electrophoretic measurements.
The unexpectedly long lifetime that occurs even for uncoated MBs has been verified by another research group \cite{ushikubo}, though it remains to be settled as to whether the stability is generic to submicron MBs or due to any artifacts such as tiny submicron dusts that could cover a part of the submicron MB-surfaces. 

The MB density was evaluated by counting the averaged total number of Brownian particles in the obtained micrographs where the absolute value of the number density has been evaluated using an SP solution of a known number density as a reference sample.
We adjusted the SP particle count to that of the MBs by diluting the SP solutions.
We thus found an MB density of approximately 3 fmol/L, which means that there are a few particles in a 100-$\mu\mathrm{m}$ cube.
This density is so low that the slow kinetics of coagulations between MB-PE complexes may be expected.
Moreover, the results in section 3.3 revealed that excess PE chains exist around MB-PE complexes that are neutralized instead of being overcharged and that the surrounding cationic PEs could prevent one complex from approaching the other similar to PE-brushes on polymer colloids.
These findings explain why isolated MB-PE complexes with an invariant diameter and submicron MBs were observable in our experiments.

\subsection{Comparison between effective and saturated charges per MB}
In general, the electric double layer that is formed by counterions, or dissociated small ions from ionic groups on each colloid, considerably reduces the actual charge compared to the effective one.
For the SPs, we obtained the ratio, $Z_{\mathrm{SP}}/Z^*_{\mathrm{SP}}\approx 4600$, between bare and effective charges using the results in sections 3.2 and 3.3.
The large difference implies that the effective charge density on each MB, which is close to that of each SP, may be far from the actual MB value.
We have an indicator of such discrepancy because the reduction of the charges that each MB effectively carries is suggested by reaching an experimentally determined saturation value;
otherwise, the bare value should be identified with the effective density unsaturated.

A collection of various experiments demonstrated that the saturation number $Z^*_x$ of effective charges satisfies the relation,
   \begin{equation}
   Z^{\mathrm{upper}}_x=k\frac{l_B}{a},
   \label{saturation}
   \end{equation}
in accordance with the theoretical prediction based on the Poisson-Boltzmann approach \cite{saturation}.
In eq. (\ref{saturation}), the coefficient $k$ is in the range of $5\leq k\leq8$ \cite{saturation}.
Equation (\ref{saturation}) provides that $3400 < Z^{\mathrm{upper}}_{\mathrm{SP}} < 5500$, whereas experiments have given $Z^*_{\mathrm{SP}}\approx 3500<<Z_{\mathrm{SP}}$ within this range (see section 3.2), which validates the above discussion.

In contrast, in the case of MBs, we substitute $2a\approx400$ nm into eq. (\ref{saturation}), yielding
   \begin{equation}
   1400 < Z^{\mathrm{upper}}_{\mathrm{MB}} < 2400.
   \label{mb-range}
   \end{equation}
The lower bound of saturated number in eq. (\ref{mb-range}) is somewhat larger than $Z^*_{\mathrm{MB}}\approx9\times10^2$.
In other words, the effective MB-charges is not a saturated value, allowing us to identify the effective charge density, $\sigma^*_{\mathrm{MB}}\approx -1.7\times10^{-3}e\,\mathrm{C}/\mathrm{nm}^2$ in section 3.2, with the actual value that MBs carry.
The present value of $-1.7\times10^{-3}e\,\mathrm{C}/\mathrm{nm}^2$ means that one small anion exists in a $30\,{\mathrm{nm}}\times 30\,{\mathrm{nm}}$ square on MB-surfaces. This quantity seems plausible because hydroxyl ions from water molecules mainly create anionic groups that are in adsorption equilibrium on the air-water interface of MBs.

\subsection{A model for microbubble-polyelectrolyte complexation}
The conventional form of the free energy functional is $F\{c_m\}=\int d{\bf r}f[c_m({\bf r})]$ for PEs in the presence of an oppositely charged wall where $c_m({\bf r})$ denotes the inhomogeneous distribution of monomer concentration.
The free energy density $f[c_m({\bf r})]$ reads in the low salt limit \cite{review}:
   \begin{eqnarray}
   \frac{f[c_m({\bf r})]}{k_BT}=
   &&\frac{c_m}{M}\ln\left(\frac{c_m}{M}\right)-\frac{c_m}{M}
   +\frac{b^2}{24}\left(\frac{|\nabla c_m|^2}{c_m}\right)\\\nonumber
   &&-\alpha c_m^s u_s-\frac{\mu_p}{M}\left(c_m-c_m^b\right),
   \label{m-energy}
   \end{eqnarray}
where $M$ denotes the degree of polymerization, $b$ the Kuhn length of used PEs, $\alpha$ the averaged fraction of attached monomers in a single polyelectrolyte, $c_m^s$ the monomer concentration at an oppositely charged surface, $-u_s$ a constant electrostatic interaction potential at the surface in the unit of $k_BT/e$, $\mu_p$ the chemical potential per polyelectrolyte chain in the $k_BT$-unit, and $c_m^b$ the bulk density of monomers.
Equation (\ref{m-energy}) states that a determining factor of the conventional polyelectrolyte systems in adsorption equilibrium is not polymer density but monomer concentration.
We have omitted segment repulsion terms due to both excluded volume and electrostatic contributions in eq. (\ref{m-energy}) for simplicity partly because these terms modify the detailed behaviors but never change the determining factor and partly because intrachain interactions can be included in the conformational entropy term that depends on various parameters such as persistence length \cite{dobrynin}.

The first three terms on the right hand side (rhs) of eq. (\ref{m-energy}) correspond to entropic contributions; the first two terms arise from the translational entropy of PE chains, whereas the third term is associated with a loss of conformational entropy.
The translational entropy terms with the $1/M$ prefactor are usually negligible in comparison to the free energy variance due to the loss of conformational entropy, if a constraint exists on chain configurations.
The fourth term on the rhs of eq. (\ref{m-energy}) is associated with the ionic parings between the ionic groups that polyelectrolytes carry and the opposite charges fixed on an interface.
Equation (\ref{m-energy}) manifests that monomer concentration, $c_m({\bf r})$, rather than polymer density, determines the attractive electrostatic interaction energy as well as the above entropic contribution.

There are conditions necessary to replace monomer concentration, used as an appropriate parameter in eq. (\ref{m-energy}), for polymer density.
Before providing these conditions, we mention non-polymeric systems, including surfactant and oligomer solutions, where a simple form of the adsorption equilibrium is given by
    \begin{equation}
    k_BT\ln \phi^b=-\Delta u,
    \label{chemical}
    \end{equation}
where $\phi^b$ denotes the bulk volume fraction of non-adsorbed molecules and $-\Delta u$ the potential difference between absorbed and desorbed molecules.
The translational entropy of freely mobile molecules yields the logarithmic term in eq. (\ref{chemical}), which is balanced by an energetic gain $\Delta u$ that is independent of the molecular size.
The Langmuir adsorption isotherm is reproduced by adding, to the rhs of eq. (\ref{chemical}), the translational entropy term of the adsorbed molecules that can still move on the surfaces \cite{diamant}.
Equation (\ref{chemical}) implies the relevant modifications of the free energy form, eq. (\ref{m-energy}), for MB systems.
One necessary condition for this modification is to regard the loss of conformational entropy as negligible in comparison to the translational entropy terms in eq. (\ref{m-energy}).
The other condition is to assume that the attractive interaction energy per single polyelectrolyte is independent of chain length.
The latter assumption means that the contact number of ionic groups on polyelectrolyte chains on oppositely charged interfaces is invariant with respect to the variance of polymerization degrees.

By focusing on the different nature of surface charges between MBs and usual polymer colloids, we discuss the underlying mechanism that supports the above conjectures on the alteration of free energy form (\ref{m-energy}).
Surface charges on polymer colloids are fixed to the cores, whereas small anions, such as hydroxyl ions, are dissociated after a temporary attachment on an MB-surface .
Our idea is to ascribe the irrelevance of conformational entropy changes to the fast adsorption-desorption kinetics of small anions on MB-surfaces based on the following dynamical discussions.
First, any constraints on PE conformations would be effectively absent when there is a large difference between characteristic times of configurational fluctuations and adsorption-desorption of the anions.
Moreover, the loss of conformational entropy is negligible, and the third term on the rhs of eq. (\ref{m-energy}) can be dropped when monomers have no difficulty in changing configurations.
Consequently, the translational entropy contribution survives as the remainder of the entropic terms.

It is thus significant for the understanding of MB systems that the attached cations of PEs are non-electrostatically dissociated from the MB-surface due to the desorption of small anions.
This feature also affects the electrostatics, leading to the relation $\alpha=\gamma/M$, with $\gamma$ being a constant that is independent of the polymerization degree $M$, as shown below.
The fourth term on the rhs of eq. (\ref{m-energy}) is then rewritten as
   \begin{equation}
   -\alpha c_{m}^su_s=-\gamma\frac{c_m^s}{M}u_s=-\gamma
   c_{p}^s u_s,
   \label{supposition}
   \end{equation}
where we have introduced the PE density $c_p({\bf r})=c_m({\bf r})/M$ and $c_{p}^s$ corresponds to the PE density at an MB-surface.
This expression indicates that the attractive electrostatic adsorption energy explicitly depends on PE density and is independent of chain length.

\begin{figure*}[tbh]
\begin{center}
\includegraphics[width=11cm]{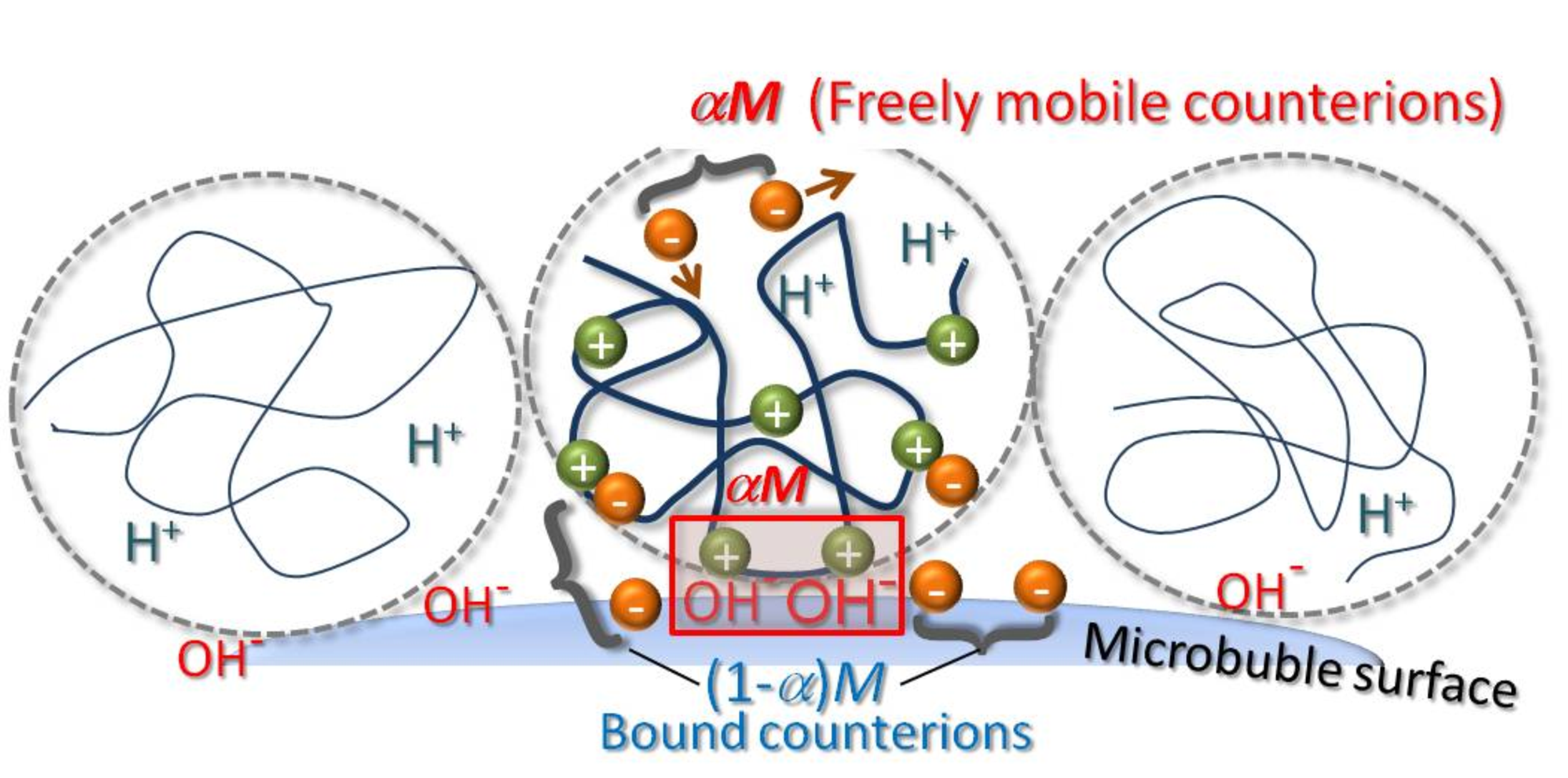}
\caption{(Color online) A schematic model in formulating the counterion free energy $A_{\mathrm{c}}$ per single polyelectrolyte. Dashed circles represent unit cells with an equal volume of $V_{\mathrm{cell}}$. While counterions with the number of $(1-\alpha)M$ are bound to either cationic groups or the air-water interface, the remaining $\alpha M$-counterions are freely mobile together with hydrogen ions because $\alpha M$-groups of cations that are paired with existing hydroxyl ions on the microbubble surface release the $\alpha M$-counterions.}
\label{f5}
\end{center}
\end{figure*}
We then prove eq. (\ref{supposition}), which represents the unusual MB electrostatics in the vicinity of oppositely charged interfaces. Emphasis is placed on the anions, instead of the partners of cationic groups on PE chains, and this allows identification of the charge cancellation on MB-surfaces with the electrostatic attachments of cationic groups to existing hydroxyl ions on the air-water interface (see also a schematic in Fig. 5).
In formulating the free energy $A_{\mathrm{c}}$ of the anions as counterions of PEs, we adopted the cell model \cite{dobrynin} that considers counterions around a single PE chain in a large unit cell (a dashed circle in Fig. 5) with size equal to the distance between chains.
Accordingly, the counterion free energy $A_{\mathrm{c}}$ per single chain is given by
    \begin{equation}
    \frac{A_{\mathrm{c}}}{k_BT}=\alpha M\ln\left(
    \alpha\overline{c}_m\right)
    -\alpha M
    +(1-\alpha)M\mu_{bc},
    \label{h-energy}
    \end{equation}
where $\overline{c}_m=M/V_{\mathrm{cell}}$ corresponds to a locally averaged density of all the counterions for an adsorbed single polyion, and $\mu_{bc}$ denotes the chemical potential of bound counterions in the $k_BT$-unit. Additionally, $\overline{c}_m=M/V_{\mathrm{cell}}$ takes a constant value within the cell of volume $V_{\mathrm{cell}}$ that is determined by a polymer density independently of the polymerization degree $M$.
While the first two terms on the rhs of eq. (\ref{h-energy}) are due to the translational entropy of counterions that are released from absorbed cationic groups and are assumed to be mobile only inside the cell, the last term includes the free energy increment due to counterions that are bound to either cationic groups or the air-water interface.

Minimization of eq. (\ref{h-energy}) with respect to $\alpha$ gives $\alpha\overline{c}_m=e^{\,\mu_{bc}}$, implying that the mean density $\alpha\overline{c}_m$ of adsorbed cationic groups is independent of surroundings.
When we also introduce a locally averaged polymer density $\overline{c}_p$ that is defined by $\overline{c}_p=\overline{c}_m/M=1/V_{\mathrm{cell}}$, the present relation, $\alpha\overline{c}_m=e^{\,\mu_{bc}}$, reads that
   \begin{equation}
   \alpha=\frac{\gamma}{M}=\frac{1}{M}
   \left(\frac{e^{\,\mu_{bc}}}{\overline{c}_p}
   \right)=\frac{V_{\mathrm{cell}}e^{\,\mu_{bc}}}{M},
   \label{d3result}
   \end{equation}
validating the relation (\ref{supposition}), or the existence of $\gamma=V_{\mathrm{cell}}e^{\,\mu_{bc}}$ independent of $M$.
Combining the above discussions, eq. (\ref{m-energy}) is reduced to a function of PE density $c_p$:
   \begin{equation}
   \frac{f[c_p({\bf r})]}{k_BT}=c_p\ln c_p-c_p
   -\gamma c_p^su_s-\mu_p(c_p-c_p^b),
   \label{reduction}
   \end{equation}
which simply consists of the translational entropy contribution of PE chains, the attractive electrostatic interaction energy of paired ions at MB-surfaces, and the chemical potential term.
The free energy form (\ref{reduction}) dependent solely on $c_p$ forms a theoretical basis that provides an explanation of the universal curve as a function not of monomer concentration but of polymer density as exhibited in Fig. 4(b).

\section{Conclusions}
While it is a challenging task to theoretically explain why the genuine MBs are stable, at least , for hours as recently reported \cite{frusawa, ushikubo},
our experiments of the long-lifetime systems have actually demonstrated that the microscopic electrophoresis method is available for the detailed investigations of electric properties that submicron uncoated bubbles as well as MB-PE complexes possess. This method thereby reveals the anomalous behaviors in overcharging MBs with added cationic polyelectrolyte.
The stoichiometry that isoelectric colloid-PE complexes have ubiquitously obeyed so far \cite{conventional, dobrynin} is actually invalid for compensating MB-charges with oppositely charged PEs.
Ample cations of PEs must be added, resulting in the delayed formation of neutralized complexes that are surrounded by a sea of cationic chains.
These findings indicate that the attractive electrostatic interactions play a secondary role in forming the first layer of oppositely charged PE chains on genuine MBs.
Furthermore, the emergence of a universal behavior becomes apparent through plotting the PE-density dependences of various mobilities.
Correspondingly, we have provided a theoretical explanation of the universality, focusing on the non-electrostatic adsorption of small anions, the  counterions of cationic PEs, on MB-surfaces.
However, herein, we provide only a sketch before developing a more elaborate treatment, which remains to be explored.

\vspace{20pt}
\section*{Acknowledgment}
We thank R. Akiyama and K. Koga for helpful discussions.


\begin{thebibliography}{9}
\bibitem{stride} E. Stride and M. Edirininghe: Soft Matter {\bf 4} (2008) 2350. 
\bibitem{duncan} P. B. Duncan and D. Needham: Langmuir {\bf 20} (2004) 2567.
\bibitem{borden} M. Borden: Soft Matter {\bf 5} (2009) 716, and references therein.
\bibitem{borden2} S. Siri and M. Borden: Bubble Sci. Eng. Technol. {\bf 1} (2009) 3. 
\bibitem{gene} C. M. H. Newman and T. Bettinger: Gene Therapy {\bf 14} (2007) 465. 
\bibitem{angew1} D. G. Shchukin and K. Kohler: Angew. Chem. Int. Ed. {\bf 44} (2005) 3310. 
\bibitem{angew2} M. Winterhalter and A. F.-P. Sonnen: Angew. Chem. Int. Ed. {\bf 45} (2006) 2500.
\bibitem{lang} I. Lentacker, B. G. De Grest, R. E. Vandenbroucke, L. Peeters, J. Demeester, S. C. De Smedt and N. N. Sanders: Langmuir {\bf 22} (2006) 7273.
\bibitem{borden3} M. A. Borden, C. F. Caskey, E. Little, R. J. Gillies and K. W. Ferrara: Langmuir {\bf 23} (2007) 9401.
\bibitem{daiguji} H. Daiguji, E. Matsuoka and S. Muto: Soft Matter {\bf 6} (2010) 1892. 
\bibitem{takahashi} M. Takahashi: J. Phys. Chem. B {\bf 109} (2005) 21858.
\bibitem{velev} K. G. Marinova, R. G. Alargova, N. D. Denkov, O. D. Velev, D. N. Petsev, I. B. Ivanov and R. P. Borwankar: Langmuir {\bf 12} (1996) 2045.
\bibitem{frusawa} H. Frusawa and M. Inoue: Chem. Lett. {\bf 40} (2011) 372.
\bibitem{palberg} T. Palberg and H. Versmold: J. Phys. Chem. {\bf 93} (1989) 5296, and references therein. 
\bibitem{huckel} B. Dunweg, V. Lobaskin, K. Seethalakshmy-Hariharan and C. Holm: J. Phys.: Condens. Matter {\bf 20} (2008) 404214; I. Pagonabarraga, B. Rotenberg and D. Frenkel: Phys. Chem. Chem. Phys. {\bf 12} (2010) 9566.
\bibitem{conventional} T. Okubo and M. Suda: J. Colloid and Interface Sci. {\bf 213} (1999) 565; A. Y. Grosberg, T. T. Nguyen and B. I. Shklovskii: Rev. Mod. Phys. {\bf 74} (2002) 329; R. Messina: J. Phys.: Condens. Matt. {\bf 21} (2009) 113102.
\bibitem{silanol} A. A. Christy and P. K. Egeberg: Analyst {\bf 130} (2005) 738, and references therein.
\bibitem{dobrynin} A. V. Dobrynin and M. Rubinstein: Prog. Polym. Sci. {\bf 30} (2005) 1049.
\bibitem{ushikubo} F. Y. Ushikubo, T. Furukawa, R. Nakagawa, M. Enari, Y. Makino, Y. Kawagoe, T. Shiina and S. Oshita: Colloids Surf. A {\bf 361} (2010) 31. 
\bibitem{saturation} P. Wette, H. J. Schope and T. Palberg:  J. Chem. Phys. {\bf 116} (2002) 10981.
\bibitem{review} R. R. Netz and D. Andelman: Phys. Rep. {\bf 380} (2003) 1.
\bibitem{diamant} H. Diamant, G. Ariel and D. Andelman: Colloids Surf. A {\bf 183-185} (2001) 259.
\end{thebibliography}
\end{document}